\documentclass[apj]{emulateapj}
\usepackage{apjfonts}
\usepackage{graphicx}
\usepackage{amsmath,amssymb}

\begin{document}

\title{Clustered Cepheid Variables 90 kiloparsec from the Galactic Center}
\author{Sukanya Chakrabarti \altaffilmark{1},
Roberto Saito \altaffilmark{2},
Alice Quillen \altaffilmark{3}, 
Felipe Gran \altaffilmark{4,5},
Christopher Klein \altaffilmark{6}  \& 
Leo Blitz \altaffilmark{6} 
}
\altaffiltext{1}
{School of Physics and Astronomy, Rochester Institute of Technology, 84 Lomb Memorial Drive, Rochester, NY 14623; chakrabarti@astro.rit.edu}
\altaffiltext{2}
{Departamento de Fisica - Universidade Federal de Sergipe, Rod. Marechal Rondon s/n - Jardim Rosa Elze, Sao Cristovao, 49.100-000, Sergipe, Brazil}
\altaffiltext{3}
{Department of Physics and Astronomy, University of Rochester, Rochester NY 14627}
\altaffiltext{4}
{Instituto de Astrof\'{i}sica, Facultad de F\'{i}sica, Pontificia Universidad Cat\'{o}lica of Chile, Av. Vicu\~{n}a Mackenna 4860, Santiago, Chile.}
\altaffiltext{5}
{Millennium Institute of Astrophysics (MAS), Santiago, Chile}
\altaffiltext{6}
{Astronomy Department, UC Berkeley, Berkeley CA 94720}

\begin{abstract}

Distant regions close to the plane of our Galaxy are largely unexplored by optical surveys as they are hidden by dust. We have used near-infrared data (that minimizes dust obscuration) from the ESO Public survey VISTA Variables of the Via Lactea (VVV) (Minniti et al. 2011; Saito et al. 2012; henceforth S12) to search for distant stars at low latitudes. We have discovered four Cepheid variables within an angular extent of one degree centered at Galactic longitude of $l = -27.4^\circ$ and Galactic latitude of $b = -1.08 ^\circ$.  We use the tightly constrained period-luminosity relationship that these pulsating stars obey (Persson et al. 2004; Matsunaga et al. 2011) to derive distances.  We infer an average distance to these Cepheid variables of 90 kpc.  The Cepheid variables are highly clustered in angle (within one degree) and in distance (the standard deviation of the distances is 12 kpc).   They are at an average distance of $\sim 2~\rm kpc$ from the plane and their maximum projected separation is $\sim 1~ \rm kpc$.  These young ($\sim$ 100 Myr old), pulsating stars (Bono et al. 2005) are unexpected at such large distances from the Galactic disk, which terminates at $\sim$ 15 kpc (Minniti et al. 2011).  The highly clustered nature in distance and angle of the Cepheid variables suggests that the stars may be associated with a dwarf galaxy, one that was earlier predicted by a dynamical analysis (Chakrabarti \& Blitz 2009).

\end{abstract}


\section{Introduction}

Studying regions close to the Galactic plane in the optical is difficult due both due to dust obscuration and source confusion.  It was only recently Feast et al. (2014) reported the discovery of five classical Cepheid variables at distances of 13 - 22 kpc from the Galactic center, towards the Galactic bulge, that may be associated with the flared atomic hydrogen disk of our Galaxy.  Two classical Cepheid variables at 11 kpc close to the plane of the Milky Way have been recently uncovered from VVV data (Dekany et al. 2015), that indicates an underlying young star cluster.  Searches for dwarf galaxies in the optical have primarily targeted high latitudes (McConnachie 2012).  The Sagittarius (Sgr) dwarf galaxy is the closest known dwarf galaxy to the plane, at a latitude of $b = -14^\circ$ (Ibata et al. 1994).  The dearth of Milky Way satellites at low latitudes (Mateo 1998; McConnachie 2012) is underscored by simulations that suggest that there may be massive, nearly dark satellites that have not yet been discovered (Boylan-Kolchin et al. 2011).  Not only dwarf galaxies, but even bright spiral galaxies are not easily seen if they are hidden behind the obscuring column of dust and gas of the Galactic disk (Kraan-Korteweg et al. 1994).

Mining data from deep infrared surveys of the Galactic plane may well uncover new dwarf galaxies and halo sub-structure.  This would alleviate several outstanding problems in near-field cosmology.  The "missing satellites problem", or the overabundance of dwarf galaxies in cosmological simulations relative to the number of observed dwarf galaxies in and around the Local Group (Klypin et al. 1999), and the "too big to fail problem", wherein there are too few massive satellites in the Milky Way relative to cosmological simulations (Boylan-Kolchin et al. 2011) are two such outstanding problems.   Yet another is the ostensibly anisotropic distribution of the Milky Way satellites (Kroupa et al. 2005).  These discepancies may be resolved by a more complete inventory of the structure of our Galaxy at low latitudes.  

We have searched for distant stars close to the Galactic plane using near-infrared data from the ESO Public survey VISTA Variables of the Via Lactea (VVV) (Minniti et al. 2011; S12), targeting the VVV disk area, which covers Galactic longitudes $-65.3^\circ < l <
-10^\circ$ within Galactic latitudes $-2.25^\circ < b < +2.25^\circ$.    The VVV survey is an ongoing 5-band photometric survey in the Z (0.87$~\mu\rm m$), Y (1.02$~\mu\rm m$), 
J (1.25 $~\mu\rm m$), H (1.64 $~\mu\rm m$) and $K_{s}$ (2.14 $~\mu\rm m$) bands (S12),
and is multi-epoch in the $K_{s}$ band, with approximately 30-40 epochs per star across the VVV disk area at the time of writing.  
In \S \ref{sec:results}, we review the methods we used to identify Cepheid variables, and present the distance and extinction values.  We discuss possible interpretations and conclude in \S \ref{sec:conclusion}.

\section{Results \& Analysis}
\label{sec:results}

The infrared photometry is from the VVV survey, which is based on aperture photometry
computed on the individual tile images (S12).
Each of the sources was observed with a median exposure time of $16$ s
per pixel, depending on the position in the tile (each exposure is $8$ s
long, and most of the area in a tile is a combination of two
pointings).  The limiting magnitude of the VVV data using aperture photometry
is $K_{s} \sim 18.0$ mag in most fields (S12).  
A particular pointing is called a ``tile'', covers $\sim1.64$ square
degree, and contains approximately $10^{6}$ stars.  
As a preliminary search, we examined the disk area of the VVV survey by applying color cuts that correspond to distant ($D > 60~\rm kpc$) red-clump
stars.  Red-clump stars have been shown to be good distance indicators (Alves 2000; Paczynski \& Stanek 1998).
Given the mean values of intrinsic near-infrared colors for red-clump stars in the Milky Way disk and the Cardelli et al. (1989) extinction law, we used the distance modulus noted in Minniti et al. (2011), which gives a color cut of $1.5<(J-K_{\rm s})<1.8$ and
$K_{s} > 17.6$ (which corresponds to distances in excess of $\sim$ 60 kpc).  Using this color cut, we saw an excess of distant red-clump stars 
at $l \sim -27^{\circ}$.   We defer a detailed analysis of the red-clump stars and other stellar populations to a future paper.

We  carried out a search for variable stars, restricting our search to faint variables, with mean $K_{s} > 15$ mag, and periods greater than three days.  We examined the variability data in five tiles close to Galactic longitude $l \sim -27^\circ$ and searched six comparison tiles at other locations in the VVV disk area.  We found four Cepheid variables at $l \sim -27^\circ$ at an average distance of $\sim$ 90 kpc,
and none in the other tiles.  The survey strategy ensures that the tiles in the VVV disk area have similar number of observations and limiting magnitude (S12).  While the control on the cadence is limited (Saito et al. 2013), we have checked that there is no significant difference in the cadence for the $l \sim -27^\circ$ tiles relative to the rest of the disk area,  i.e., the region at $l \sim -27^\circ$ is not unique in terms of the way it was observed.

In identifying Cepheids, we employed
several successive tests.  The first two tests are based on the statistical significance of the highest peak in the Lomb-Scargle periodogram, and the uncertainty of the period, respectively.  These two tests ensure that we have identified sources of a given pulsation period, and that there is a small uncertainty in the period that we derive.  For the final cut, we quantitatively assess the shapes of the light curves by calculating the Fourier parameters of the sources, as well as the skewness and acuteness parameters, and visually inspect the light curves.  

A given tile is searched using the Lomb-Scargle algorithm (Lomb 1976; Scargle 1982), and periodograms
are constructed for every source.   The statistical significance of the amplitude of the largest peak in the periodogram (Scargle 1982)
corresponds to a false alarm probability $p_{0}$.  Claiming the detection of the signal if the amplitude
exceeds the threshold value, one can expect to be wrong a fraction $p_{0}$ of the time.  Alternately, the statistical significance
level of the detection is $1-p_{0}$, and this quantity is listed in Table 1.
For the first test, we require $\bar{K_{s}} >$ 15 mag
to search for faint variables, and set the minimum and maximum period range in our variability search
between 3 - 50 days.  To pass the first test, sources have to satisfy the following conditions:
1) the period corresponding to the maximum in the Lomb-Scargle periodogram is greater
than three days, 2) the maximum in the Lomb-Scargle periodogram exceeds 90-th pericentile 
for the significance level, 3) if there are other maxima in the periodogram that are at 90-th percentile
or higher, the periods corresponding to these maxima must differ by a factor of two or less.  
The last condition amounts to requiring a clean periodogram without spuriously large multiple peaks.

In the second test, we assess the quality of the light curves of the variables that pass the tests above with a
parametric bootstrap.  Assuming a Gaussian distribution of errors, we sample the distribution one thousand times
to derive the distribution of periods for each source, which is similar to prior work (Klein et al. 2012; Klein et al. 2014)
on RR Lyrae stars.
If the mean of the period distribution agrees to within 20 \% of the period calculated from the raw data, and 
if the mean of the period distribution $\pm$ the standard deviation still exceeds three days, we consider the
period distribution to be sufficiently well constrained.   The goal of this second cut is to select sources that have a small uncertainty in the derived period, given the photometric errors.  
The Lomb-Scargle algorithm allows us to derive the period and its statistical signficance, but not the uncertainty in the derived period.   
If the width of the histogram that gives the distribution of periods from the bootstrap calculation is narrow, the uncertainty in the derived period is low.  For the sources that pass the above tests, we fit the light curves with a Fourier series (Kovacs \& Kupi 2007).
 The Fourier parameters are similar to the light curves of classical Cepheids for $P \sim 3 - 15$ days observed in the K-band (Persson et al. 2004; Bhardwaj et al. 2015).  The light curves of Cepheid variables in the optical are different in shape and amplitude from the light curves of Cepheid variables in the K-band (Matsunaga et al. 2013; Bhardwaj et al. 2015), and our comparison here is to the observed light curves of classical Cepheid variables in the K-band.  The Cepheids we list here pass all of the automated and visual checks.  Because of this multi-tiered, conservative approach, the light curve analysis is time consuming, but allows us to derive accurate distances.  It is worth noting that in addition to lower extinctions in the infrared relative to the optical, another advantage of infrared photometry of Cepheids is that it is minimally affected by metallicity variations (Bono et al. 2010; Freedman et al. 2010).

\begin{figure}
\begin{center}
\includegraphics[scale=0.5]{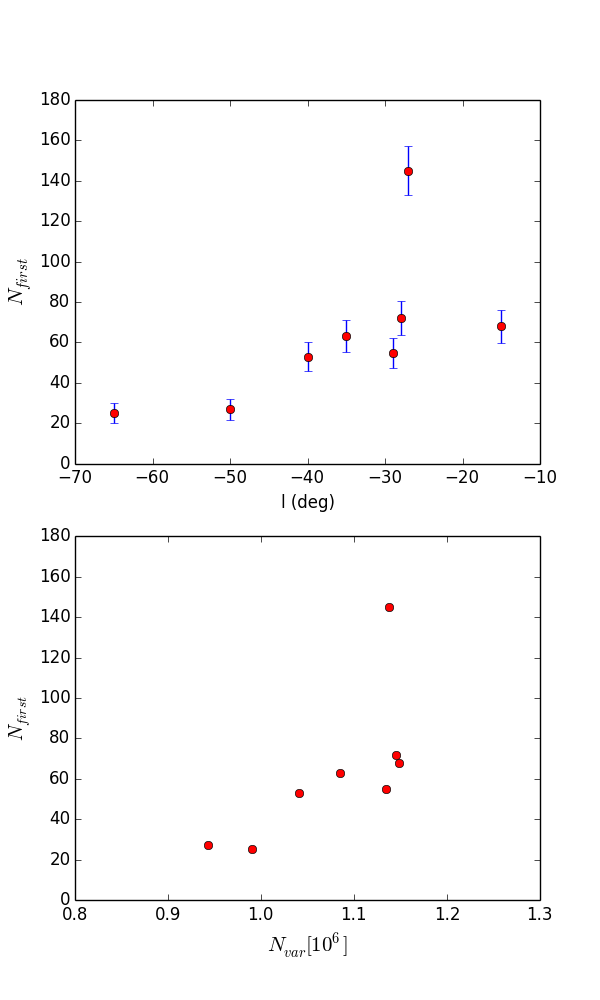}
\caption{The number of sources that pass the first of our tests (requiring statistical significance greater than 90-th percentile, $P >$ 3 days, $K_{s} > 15~\rm mag$) is shown as a function of longitude in the top panel, and as a function of the total number of variable stars in the tile (normalized to one million) in the bottom panel.  The error bars in the top-panel is from Poisson noise. 
\label{f:comb}}
\end{center}
\end{figure}

The tiles close to longitude $l \sim -27^\circ$ produce a significantly larger number of variables
that pass the first of our tests than the other six tiles we examined (at $l = -15^{\circ},-29^{\circ}, -35^{\circ}, -40^{\circ}, -50^{\circ}, -65^{\circ}$).  Figure 2 of S12 depicts the VVV survey area.  The number of sources in tiles d027, d065, d103, and d141 
that are centered at $l \sim -27 ^\circ$ and extend upwards in latitude (S12), produce $\sim 100-200$ sources
that pass our first cut.  In contrast, the average number that pass the first cut from the comparison tiles is $\sim$ 60.  If we consider 
this background number to be the mean of a Poisson distribution, and randomly sample a Poisson distribution with this mean value, values in excess of 100 are above 5-$\sigma$, i.e., they are statistically extremely unlikely to occur by chance.  Figure \ref{f:comb} shows the number of sources that pass the first of our tests as a function of longitude (the value at $l \sim -27 ^\circ$ is an average over latitude), as well as a function of the total number of variable stars in the tile.  While the number of sources that pass the first of our tests has some correlation with the number of variable stars (which is not unexpected), the region at $l \sim -27 ^\circ$ is a clear outlier. In some tiles centered at $l \sim -27 ^\circ$, there were a significant number that passed our automated and visual analysis of the light curves, but did not pass our visual inspection of the images, due to the possibility of spikes or blending.  
The number of Cepheid variables that we report here from the final cut is very likely an underestimate.  

The phase folded light curves of the Cepheid variables, which show a clear resemblance to each other, and corresponding images are shown in Figure 1.   Table 1 summarizes the derived distances and other parameters for the Cepheids.  To estimate the dust extinction from the excess color, we use the quasi-simultaneous
single epoch VVV measurements in the $J$, $H$ and $K_{s}$ bands ($\sim$ 190 s between each band).  The near-infrared amplitudes of 
classical Cepheids are relatively small (Persson et al. 2004) and as such an estimate of the
dust extinction from the single epoch measurements of the colour should be sufficient.
The average extinction-corrected $(J-K_{s})$ colours of the Cepheids is $\sim$ 0.4, which is
consistent with the colors of short-period classical Cepheids (Persson et al. 2004) in the LMC.

\begin{table*}
\centering
        \caption{Data For Individual Cepheids and Derived Parameters}

          \begin{tabular}{@{}lccccccc@{}}
          \hline

VVV ID  &       $l~\rm (deg)$      &    $b~\rm (deg)$        &    D (kpc)  &  P (day)       &  $\bar{K_{s}}$       &  Significance Level  \\
\hline

VVVJ162559.36-522234.0     &   -27.5971       &  -2.23686           &  92         &  3.42        & 16.04            &   91 \%           \\
VVV J162328.18-513230.4        &   -27.2729    &  -1.37557          &  100      &  4.19      &   16.12    &         93 \% \\
VVVJ162119.39-520233.3      &  -27.8621       & -1.49382            &  71        &   5.69        & 15.1              &    97 \%           \\
VVV J161542.47-494439.0       &  -26.8882    &     0.768427         &  93        &  13.9        & 15.6      &         98 \%    \\

\hline

\end{tabular} 

\small {VVV ID, Galactic longitude ($l$) and latitude ($b$), $D$ is the distance from the sun, $P$ is the pulsation period, $\bar{K_{s}}$ is the mean $K_{s}$-band magnitude, the last column is the significance level of the highest peak in the Lomb-Scargle periodogram.  }

\end{table*} 

\begin{figure}
\begin{center}
\includegraphics[scale=0.65]{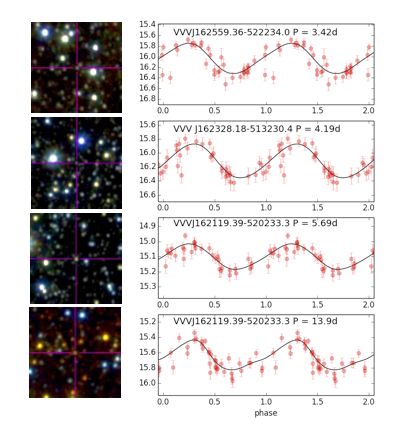}
\caption{$JHK_{\rm s}$ false color
image of the Cepheid variables, with phase folded $K_{\rm s}$-band
light curves. All fields are $30"\times30"$, oriented in Galactic
coordinates. The VVV ID and period are also listed in Table 1, along
with the Galactic latitude, longitude, distances and average $K_{\rm
s}$-band magnitude.  The four light curves have a clear resemblance to each other, and a quantitative assessment of their shapes shows they are similar to those of $K_{s}$-band light curves of classical Cepheids.
\label{f:cep}}
\end{center}
\end{figure}

We adopt the period-luminosity relations of classical Cepheids in the LMC (Matsunaga et al. 2011), with a LMC distance modulus
of 18.5 mag and interstellar extinction value of $A_{K_{s}} = 0.02$ mag for the LMC
direction.  This gives the distance modulus $\mu$ for a Cepheid with pulsation
period $P$ (Feast et al. 2014):

\begin{equation}
\mu= K_{s} - A_{K_{s}} + 3.284~\rm log(P) + 2.383 \; ,
\end{equation}
where $A_{Ks}$ is the extinction in the $K_{s}$ band, which we can express in terms
of the colour excess:

\begin{equation}
A_{Ks} = 0.6822 E(J - K_{s}) \; ,
\end{equation}
where $E(J-K_{s}) = (M_{J} - M_{K_{s}})_{\rm obs} - (M_{J} - M_{K_{s}})_{\rm int} $ is the difference between the observed and intrinsic colors,
and we adopt the period-luminosity relations of classical Cepheids in the LMC (Matsunaga et al. 2011) and the Cardelli et al. (1989) extinction law.  The single-epoch colors and extinctions in the $K_{s}$ and $J$ bands,
along with the extinction corrected colors are listed in Table 2. Using extinction values from dust maps derived
from far-IR colors (Schlegel et al. 1998) leads to slightly larger values close to the plane of the Galaxy along these lines of sight, as well as recent work that is based on spectra from the Sloan Digital Sky Survey (Schlafly \& Finkbeiner 2011).  If we consider the standard deviation of these three values to be the uncertainty in the dust extinction, and include the photometric errors and uncertainty in the period distribution to derive the uncertainties in the distance, on average this gives a distance uncertainty of $\sim$ 20 \%, where the dominant term is the uncertainty in the extinction.  The dust extinction in this area (Schlegel et al. 1998) is not unusual, i.e., this area is neither a high or low region of dust extinction.  

\begin{table*}
\centering
        \caption{Photometry Data And Extinction}

%\caption*{

        \begin{tabular}{@{}lccccccc@{}}
          \hline

VVV ID                                   &   J            &     H          & $K_{s}$    & $A_{K_{s}} $   &   $A_{J}$      &  $(J-K_{s})_{\rm corr}$    \\ 
\hline

VVVJ162559.36-522234.0    & 17.078   &  16.429  & 16.175    & 0.348          & 0.83          & 0.42  \\
VVV J162328.18-513230.4     &    17.88  &   17.09      & 16.7         &   0.53          &   1.25          &       0.44        \\
VVVJ162119.39-520233.3    & 16.416  & 15.414   & 14.96     & 0.7         &  1.68          &  0.48    \\
VVV J161542.47-494439.0    &    18.71   &  16.69      &    15.45     &  1.89            &   4.52            &   0.6              \\

\hline

\end{tabular}
\small {\center{Single epoch VVV photometry in the $J, H$ and $K_{s}$ bands, with extinction values derived from the color excess assuming the Cardelli et al. (1989) extinction law, along with the extinction corrected $(J-K_{s})$ color.}
}

\end{table*}

\begin{figure}
\begin{center}
\includegraphics[scale=0.35]{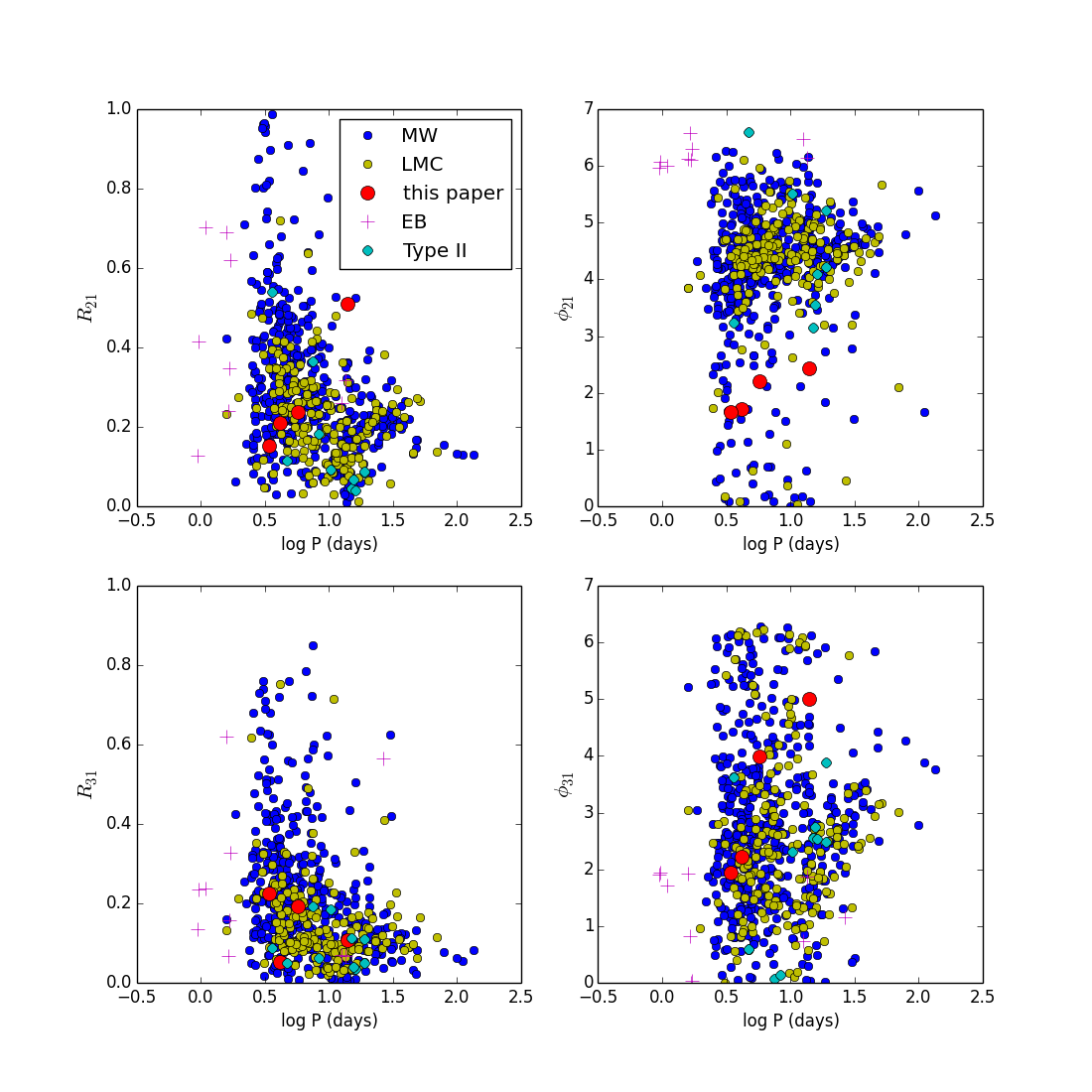}
\caption{The Fourier parameters (defined by Eqn 3 and corresponding text) are plotted vs period.  We have plotted here the Fourier parameters derived from K-band light curves of Type I, classical Cepheids in the Milky Way (marked "MW"), Type I classical Cepheids from the LMC (marked "LMC") (Bhardwaj et al. 2015), the Cepheid variables discovered at $\sim$ 90 kpc (marked "this paper"), eclipsing binaries (Matsunaga et al. 2013) (marked "EB"), and Type II Cepheids  (Matsunaga et al. 2013) (marked "Type II").
\label{f:fourier}}
\end{center}
\end{figure}

Short-period ($\sim$ day) close contact eclipsing binaries like W Ursa Majoris stars can mimic the sinusoidal light curves of RR Lyrae stars, which have periods of $\sim$ day (Rucinski 1993), but is less of a concern for long-period variables.  To ensure that the shapes of the light curves are quantitatively similar to classical Cepheids, we compute their Fourier parameters, as well as the skewness and acuteness parameter, and visually inspect all the light curves.
We have computed the Fourier parameters of our sources (Figure \ref{f:fourier}), following Kovacs \& Kupi (2007).  Out to fourth-order the Fourier series is expressed as:

\begin{equation}
m(t) = A_{0} + \sum_{i=1}^{4} A_{i} \rm cos\left(2\pi i t/P + \phi_{i}  \right)\; ,
\end{equation}
where $m(t)$ is the light curve, $P$ is the period, and $A_{i}$ and $\phi_{i}$ are the amplitudes and phases respectively.
The top panel shows $R_{21} = A_{2}/A_{1}$ and $\phi_{21} = \phi_{2} - 2 \phi_{1}$, and the bottom panel shows $R_{31} = A_{3}/A_{1}$ and $\phi_{31} = \phi_{3} - 3 \phi_{1}$.  Eclipsing binaries have $\phi_{21}$ and $\phi_{31}$ values close to 2$\pi$ or zero, which reflects their symmetric variations (Matsunaga et al. 2013).  Thus, the Fourier parameters of our sources indicate that they are not eclipsing binaries.   Bhardwaj et al. (2015) provides a compilation of the Fourier parameters of a large number of Galactic and LMC Type I Cepheids across a range of wavelengths.  This work shows the differences in the shape of the light curve in the K-band relative to the I-band, as well as differences between the K-band light curves of Cepheids in the Galaxy and in the LMC.   We have overplotted in Figure \ref{f:fourier} the Fourier parameters of Type II Cepheids in the Milky Way (Matsunaga et al. 2013).   These Type II Cepheids tend to have lower $\phi_{31}$ values than classical, Type I Cepheids, but there is scatter in the Fourier parameters derived from K-band light curves.

The shape of the light curve can be further quantified by the skewness ($S_{k}$) and acuteness ($A_{c}$) parameters:
\begin{equation}
S_{k}=\frac{1}{\phi_{\rm rb}} -1, ~~~~ \phi_{\rm rb} = \phi_{\rm max}-\phi_{\rm min},~~~~ A_{c} = \frac{1} {\phi_{\rm fw}} -1 \; ,
\end{equation}
where $\phi_{\rm max}$ and $\phi_{\rm min}$ are the phases corresponding to the minimum and maximum of the rising branch, and $\phi_{\rm rb}$ is therefore the phase duration of the rising branch, and $\phi_{\rm fw}$ is the full width at half maximum of the light curve.  Bhardwaj et al. (2015) demonstrated that the skewness parameter derived from I-band light curves is significantly higher than K-band light curves.  The average skewness parameter of our sources is $\sim$ 0.63 and the average acuteness parameter is $\sim$ 0.8, which is comparable to classical Cepheids observed in the K-band (Bhardwaj et al. 2015). 

\section{Discussion \& Conclusion}
\label{sec:conclusion}

By employing a series of successive tests to determine the periods of variable stars, the uncertainty in their periods, and a quantitative assessment of the light curve shape, we have found four Cepheid variables within an angular extent of one degree centered at Galactic longitude of $l = -27.4^\circ$ and Galactic latitude of $b = -1.08 ^\circ$, at an average distance of 90 kpc.  These successive tests are not satisfied at any of the other locations where we searched for Cepheid variables.  Spectroscopic observations would be useful to confirm the spectral type and determine a radial velocity.  Type II Cepheids that are part of the Galactic halo are not expected to be clustered within a degree, which is what we see here.  
Type II Cepheids that are part of a dwarf galaxy can be clustered. There are many more Type I, classical Cepheids than Type II Cepheids; the OGLE survey has detected 3361 Type I, classical Cepheids in the LMC, and 197 Type II Cepheids (Soszynski et al. 2008; Soszynki et al. 2008a).  Unless this object is as massive and extended as the LMC, one would expect that these sources are more likely to be Type I rather than Type II Cepheids.  If they are Type II Cepheids, they would be at an average distance of $\sim$ 50 kpc (Matsunaga et al. 2013), and such a concentration of Type II Cepheids (which are very rare) is unexpected beyond the edge of the Galactic disk.   Therefore, on the basis of the Fourier parameters, skewness and acuteness parameters, and their angular concentration, we conclude these sources are Type I Cepheids.

Earlier work (Chakrabarti \& Blitz 2009) predicted that the observed perturbations in the atomic hydrogen disk of our Galaxy (Levine, Blitz \& Heiles 2006)
are due to a recent (300 Myr years ago) interaction with a dwarf satellite galaxy that is one-hundredth the mass of our Galaxy,
currently at a distance of 90 kpc from the Galactic center, close to the plane, and within
Galactic longitudes of $-50^{\circ} < l < -10^{\circ}$ (Chakrabarti \& Blitz 2011).  
This methodology was applied to spiral galaxies with known, tidally dominant optical companions to provide a proof of principle of the method (Chakrabarti et al. 2011).   

There are no known dwarf galaxies that have
tidal debris at this location.  
The tidal debris of the Sgr dwarf does not not extend to within $\sim$
twenty-five degrees of Galactic longitude of $l=-27^\circ$ (Carlin et al. 2012), and the Magellanic stream does not extend to within $\sim$ 40 degrees of this region (Putman et al. 2003).  
%Expectations from simulations of the Sgr dwarf (Purcell et al. 2011)
%also do not indicate that tidal debris should be expected in this longitude and radial range.  
The Canis Major overdensity was identified as an excess of M-giant stars from the Two Micron All Sky Survey (2MASS)
at $(l,b) = (-120^\circ,-8^\circ)$, at a distance of $\sim$ 7 kpc from the Sun (Martin et al. 2004).
Its proximity to the Milky Way indicates that this overdensity is also unlikely to be associated with the
Cepheids we report here.

These are the most distant Cepheid variables close to the plane of our Galaxy discovered to date.  The fact that the Cepheids that we detect are at an average distance of 90 kpc, highly clustered in angle (within one degree) and in distance (within 20 \% of the mean value of 90 kpc), is difficult to explain without invoking the hypothesis of these stars being associated with a dwarf galaxy, which may be more extended in latitude than can be determined from the VVV survey alone.  Constraining the structure of this object should be possible with future deeper observations.

\bigskip
\bigskip

\acknowledgements
We gratefully acknowledge use of data from the ESO Public Survey
programme ID 179.B-2002 taken with the VISTA Telescope, and data
products from the Cambridge Astronomical Survey Unit.  R.K.S.
acknowledges support from CNPq/Brazil through projects 310636/2013-2
and 481468/2013-7.  F.G. acknowledges support from CONICYT-PCHA Mag\'{i}ster National 2014-22141509.
S.C. thanks M. Feast, B. Madore, C.F. McKee, J. Scargle, and G. Kovacs for helpful discussions.

\end{document}